\RequirePackage{fix-cm}
\documentclass[smallcondensed]{svjour3}     
\smartqed  
\usepackage{graphicx}
\usepackage{mathptmx}      
\makeatletter
\def\cl@chapter{\@elt {theorem}} 
\makeatother
\usepackage{mathrsfs,amsmath}
\usepackage{booktabs}
\usepackage{multirow}
\usepackage{array}
\newcolumntype{C}[1]{>{\centering\arraybackslash} m{#1}}
\usepackage[capitalise]{cleveref} 
\usepackage{gensymb} 
\usepackage{amssymb} 
\usepackage{tabularx}
\usepackage{subcaption} 
\usepackage[table]{xcolor} 
\journalname{International Journal of Computer Assisted Radiology and Surgery}
\begin{document}

\title{Generation of annotated multimodal ground truth datasets for abdominal medical image registration}

\titlerunning{Generation of Multimodal Registration Ground Truth}        

\author{Dominik F. Bauer$^1$ \and
        Tom Russ$^1$ \and
        Barbara I. Waldkirch$^1$ \and
        Christian Tönnes$^1$ \and
        William P. Segars$^2$ \and
		Lothar R. Schad$^1$ \and
		Frank G. Zöllner$^{1*}$ \and
		Alena-Kathrin Golla$^{1*}$
}

\authorrunning{Bauer et al.} 

\institute{Dominik Bauer\\
\email{dominik.bauer@medma.uni-heidelberg.de}\\
ORCiD: 0000-0002-8781-2882\\
Tel.: +49-621-3836551\\
\newline
$^1$ Computer Assisted Clinical Medicine, Mannheim Institute for Intelligent Systems in Medicine, Medical Faculty Mannheim, Heidelberg University, Germany\\
$^2$ Carl E. Ravin Advanced Imaging Labs, Department of Radiology, Duke University, United States
\newline
* These two authors share senior authorship.
}
\date{Received: date / Accepted: date}

\maketitle

\begin{abstract}
\textbf{Purpose} Sparsity of annotated data is a major limitation in medical image processing tasks such as registration. Registered multimodal image data are essential for the diagnosis of medical conditions and the success of interventional medical procedures. To overcome the shortage of data, we present a method that allows the generation of annotated multimodal 4D datasets.\\
\textbf{Methods} We use a CycleGAN network architecture to generate multimodal synthetic data from the 4D extended cardiac‐torso (XCAT) phantom and real patient data.
Organ masks are provided by the XCAT phantom, therefore the generated dataset can serve as ground truth for image segmentation and registration. Realistic simulation of respiration and heartbeat is possible within the XCAT framework. To underline the usability as a registration ground truth, a proof of principle registration is performed.\\
\textbf{Results} Compared to real patient data, the synthetic data showed good agreement regarding the image voxel intensity distribution and the noise characteristics. The generated T1-weighted magnetic resonance imaging (MRI), computed tomography (CT), and cone beam CT (CBCT) images are inherently co-registered. Thus, the synthetic dataset allowed us to optimize registration parameters of a multimodal non-rigid registration, utilizing liver organ masks for evaluation.\\
\textbf{Conclusion} Our proposed framework provides not only annotated but also multimodal synthetic data which can serve as a ground truth for various tasks in medical imaging processing. We demonstrated the applicability of synthetic data for the development of multimodal medical image registration algorithms.
\end{abstract}

\section{Introduction}

\subsection{Background}

Multimodal imaging plays an important part in the diagnosis of cancers, such as liver cancer \cite{tempany2015multimodal}. A variety of treatment options are available for hepatocellular carcinoma (HCC), the sixth most common malignancy worldwide and the third leading cause of cancer-related deaths \cite{memon2011radioembolization}. These include interventional procedures such as transarterial chemoembolizations (TACE) \cite{waldkirch2019multimodal} or radioembolization \cite{spahr2019multimodal}. The treatment planning benefits from using multimodal registration to combine pre- and intrainterventional data. Each imaging modality has strengths and weaknesses. Image registration enables the fusion of complementary information of each modality.

The lack of convenient ground truth data is a major limitation in the field of medical image segmentation and registration \cite{zollner2020image}. The generation of organ masks for segmentation requires labor-intensive manual annotation. For the development of image registration algorithms (especially for non-rigid image registration methods) and validation of registration accuracy, the ground truth is generally not available \cite{zollner2020image}. This is because the patient positioning in-between scans usually cannot be reproduced, particularly in the case of multimodal imaging. In the abdomen, the variable content of the bladder and bowel and additional patient motion like respiration and heartbeat further exacerbate the problem.

\subsection{Related Work}

To evaluate registration results or to train deep learning registration approaches, either anatomical multi-label organ masks or landmarks are required. However, generating these labels is labor-intensive, subjective or even impractical for large datasets. Pluim \textit{et al.} introduced a semi-automatic framework that allows the creation of a large number of high-quality landmarks with minimal user effort \cite{pluim2016truth}. Established ground truth datasets for the validation of image registration are usually only available for brain imaging. For instance, the simulated brain database (BrainWeb) provides simulated MRI imaging sequences (T1-weighted, T2-weighted, and proton density) \cite{cocosco97brainweb}. The images are perfectly aligned, since they are calculated from the same model.

Image synthesis is able to reduce multimodal registration problems to monomodal problems by first converting one modality into the other. Modality reduction has shown improvements in registration accuracy for the brain \cite{roy2014mr} and the pelvis \cite{cao2017dual}.

For MRI-only radiotherapy planning Wolterink \textit{et al.} demonstrated feasible results using a CycleGAN approach for MRI-to-CT translation and showed that training with unpaired images is superior to training with paired images \cite{Wolterink2017}.

Using the digital 4D extended cardiac‐torso (XCAT) phantom \cite{Segars2010} instead of patient images for image synthesis is beneficial, because organ masks and motion displacement fields are provided by the phantom.
Tmenova \textit{et al.} presented a CycleGAN to synthesize X-ray angiograms from the XCAT phantom, which proved to be useful as a data augmentation strategy \cite{tmenova2019cyclegan}.
Russ \textit{et al.} synthesized abdominal CT images using a CycleGAN and the XCAT phantom \cite{russ2019synthesis}. They showed that a vessel segmentation network trained on a combination of real CT and synthetic CT images achieved a superior performance compared to a network trained only on real data.
Analytical models \cite{paganelli2017tool,wissmann2014mrxcat} and a GAN approach \cite{abbasi20204d} to transform the CT XCAT phantom into cardiac or abdominal MRI images have been developed.
To our knowledge, no multimodal registration ground truth dataset of the abdomen created from the same XCAT or digital phantom has been reported in the literature.

\subsection{Contribution}

Our approach to bypass the lack of ground truth data in image registration and segmentation is the generation of a mulitmodal synthetic dataset from the XCAT phantom. In this work, we focus on multimodal image registration. The synthesis is performed via CycleGAN networks, which were seperately trained for each modality. To improve the preservation of high-contrast structures, we extend the CycleGAN generator loss with an intensity loss and a gradient difference loss.

Interventions are often monitored via cone beam CT (CBCT), whereas CT and MRI images are taken for diagnosis beforehand to assist the navigation during intervention \cite{waldkirch2019multimodal}. Thus, our multimodal dataset consists of T1-weighted MRI, CT, and CBCT images. We use XCAT data in the inhaled and exhaled motion state to synthesize images. Since the same XCAT phantom is used as the starting point for all modalities, the resulting multimodal synthetic data is perfectly co-registered. Displacement fields for respiratory movements and segmentation masks for all organs are provided by the XCAT phantom. Therefore, it serves as a ground truth dataset for registration. 

To demonstrate the utility of the multimodal dataset for the optimization of registration algorithms, we evaluate a multimodal non-rigid registration for varying parameter settings. We focus on the registration of the liver, however, the registration quality can be assessed for any other organ.

\section{Materials and Methods}

\subsection{Image Synthesis Framework}

\begin{figure}[ht]
\centering
	\includegraphics[width=0.8\textwidth]{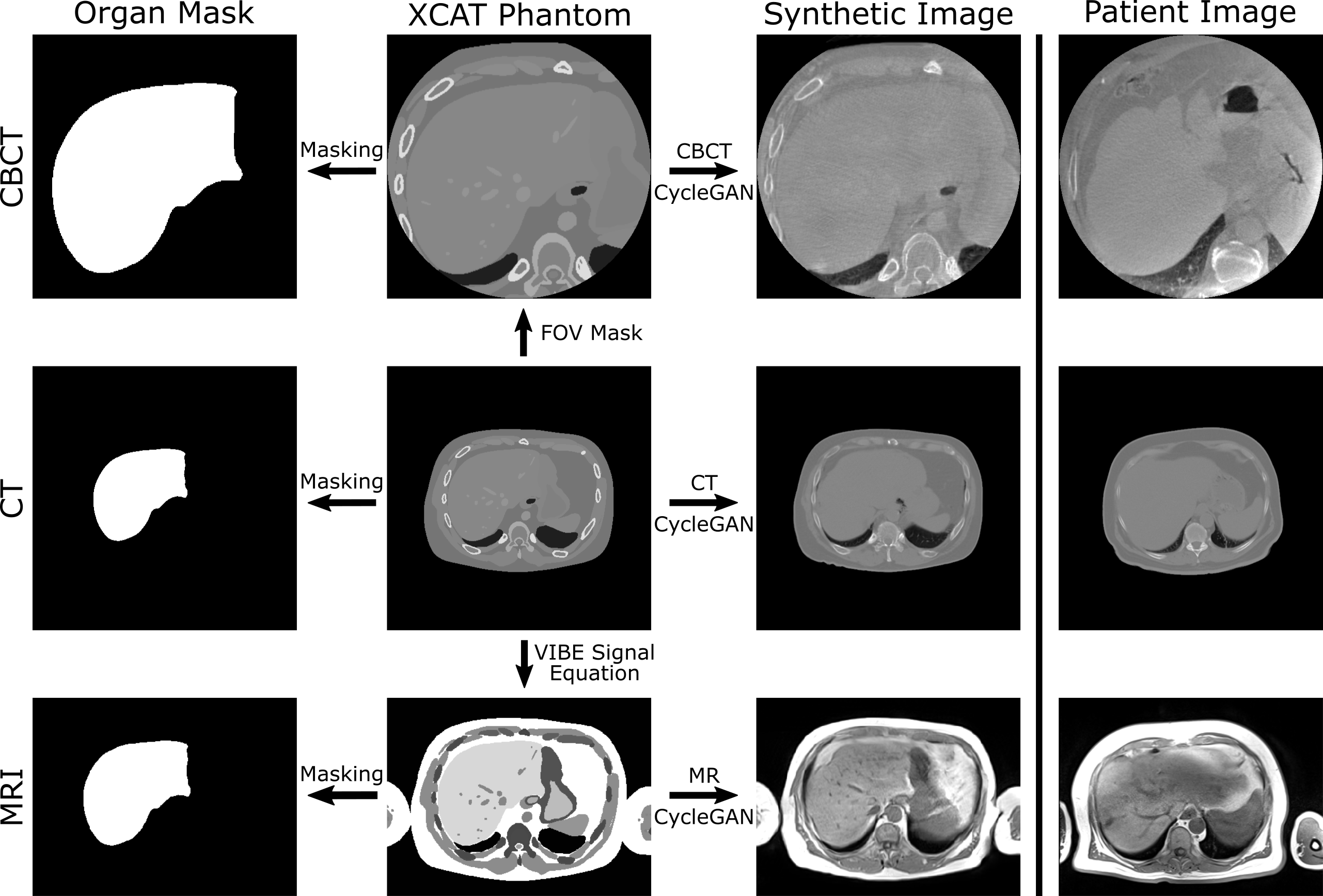}
	\caption{Schematic of the simulation framework. Starting point is the CT XCAT, from which CBCT and MRI versions are derived. Synthetic CT, CBCT, and MRI images are created via separately trained CycleGAN networks. Organ masks can be obtained from the XCAT phantoms. Patient images that are used to train the CycleGANs are shown on the right hand side.}
	\label{fig:synthetic_result}
\end{figure}

\begin{figure}[h]
\centering
	\includegraphics[width=0.7\textwidth]{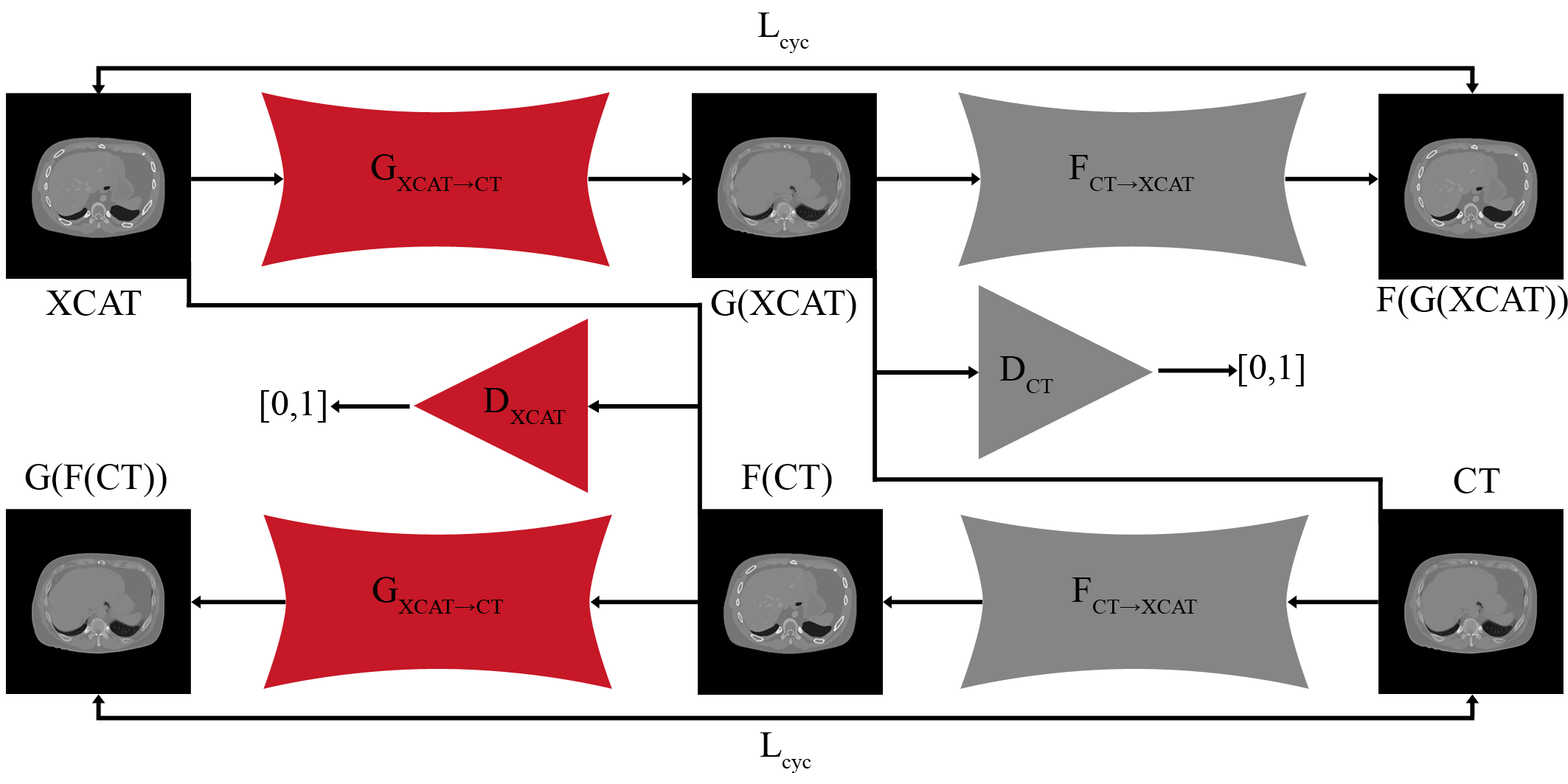}
	\caption{CycleGAN network architecture: The generators $G_{\mathrm{XCAT\rightarrow CT}}$ and $F_{\mathrm{CT\rightarrow XCAT}}$ map images from the XCAT domain to the CT domain and vice versa. CycleGAN networks for MRI and CBCT images are trained analogously.}
	\label{fig:cycleGan_Network}
\end{figure}

A schematic of our simulation framework is shown in Fig. \ref{fig:synthetic_result}. Starting from the CT XCAT phantom, CBCT and MRI XCAT versions are generated by applying a FOV Mask or by simulating the volume interpolated breathold exam (VIBE) signal equation \cite{paganelli2017tool}, respectively. Organ masks for each modality are extracted from the phantoms. Images are synthesized via CycleGAN networks using the XCAT phantom as input. CycleGANs learn the mapping between two domains $X$ and $Y$ given unpaired training samples $x\in X$ and $y\in Y$ \cite{zhu2017unpaired}. The mapping functions $G: X \rightarrow Y$ and $F: Y \rightarrow X$ are called generators. Two discriminators $D_\mathrm{X}$ and $D_\mathrm{Y}$ aim to distinguish between real images and generated images. Fig. \ref{fig:cycleGan_Network} shows the complete CycleGAN network architecture for the XCAT and CT image domain. Synthetic CT, CBCT, and MRI images are created via separately trained CycleGAN networks. The cycle consistency loss $L_{\mathrm{cyc}}(G,F)$ enforces forward and backward consistency for the generators, i.e. $F(G(x))\approx x$ and $G(F(y))\approx y$. With a least squares generative adversarial loss $L_{\mathrm{adv}}(G,F,D_{\mathrm{X}},D_{\mathrm{Y}})$, the generators are trained to generate images that cannot be distinguished from real images by the discriminator. The discriminators are 70 x 70 PatchGANs, which are trained with a least squares generative adversarial loss function. For the generators we use a Res-Net architecture with an encoding stage, 9 residual blocks and a decoding stage.

\subsection{Training and Loss Functions} 

Training is performed using the Adam optimizer with a learning rate of 0.0002. The network is trained with 256x256 pixel image patches and a batch size of 4. For each axial slice one random patch is extracted. We train the networks for 150.000 steps.

To enhance the preservation of high-contrast structures, we extend the generator loss with an intensity loss and a gradient difference loss:

\begin{align}
    L_{\mathrm{int}}(G,F)= ||(G(x)-x)||_1 + ||(F(y)-y)||_1,
\end{align}

\begin{align}
    \begin{aligned}
        L_{\mathrm{gdl}}(G, x) = \sum_{i,j}  \; & ||x_{i,j} - x_{i-1,j}|
        - \;   |G(x)_{i,j} - G(x)_{i-1,j}||^2 \\
        + \; & ||x_{i,j} - x_{i,j-1}|
        - \;   |G(x)_{i,j} - G(x)_{i,j-1}||^2.
    \end{aligned}
\end{align}

The intensity loss preserves the signal intensity of the organs provided by the XCAT phantom. As shown by Nie \textit{et al.}, the gradient difference loss prevents blurring and therefore sharpens the synthesized images \cite{nie2018medical}. The total generator loss is a combination of the previously defined losses with different weights:

\begin{align}
    \begin{aligned}
        L_{\mathrm{gen}}(G, F, D_{\mathrm{X}}, D_{\mathrm{Y}})&=
        L_{\mathrm{adv}}(G,F,D_{\mathrm{X}},D_{\mathrm{Y}})
        + \; \lambda_{\mathrm{cyc}} \;  L_{\mathrm{cyc}}(G,F) \\
        &+ \; \lambda_{\mathrm{int}} \; L_{\mathrm{int}}(G,F)
        + \; \lambda_{\mathrm{gdl}} \; (L_{\mathrm{gdl}}(G,x) + L_{\mathrm{gdl}}(F,y)).
    \end{aligned}
\end{align}

We train three CycleGAN networks for CBCT, CT and MRI with empirically chosen combinations of weights $\lambda_{\mathrm{cyc}}/\lambda_{\mathrm{int}}/\lambda_{\mathrm{gdl}}$, of 10/10/5, 10/10/5 and 10/0.4/0.4, respectively. As shown in our previous study, a combination of the gradient loss and the intensity loss yields the best results \cite{bauer2019synthesis}. Further increasing the weighting factors leads to excessive regularization and thus the networks learns an identity mapping. For the MRI networks lower over-regularization thresholds are found.

\subsection{Data}

We train our CycleGAN network to map between XCAT phantom data and real patient data. The goal is to obtain networks that generate realistic looking synthetic data using the XCAT phantoms as input. In the following paragraphs we will address the real patient and the XCAT training data separately.

\subsubsection{Patient Data}
The patient training data is retrospectively extracted from our Picture Archiving and Communication System (PACS). CT and T1-weighted MRI scans are acquired as part of routine clinical practice before TACE patients undergo a CBCT-guided TACE intervention. All scans are acquired on whole body clinical devices (Siemens Healthineers, Forchheim, Germany; CT: Somatom Emotion 16; CBCT: Artis Zeego; MRI: Magnetom Tim Trio). The MRI images are acquired at 3 Tesla with the VIBE sequence. For each modality the patient images are resampled to a unified voxel spacing given in Table \ref{table:ds_stats}. All scans include the whole liver and the narrow field of view of the CBCT scan is focused on the liver. The MRI images include arms, the CT and CBCT images do not. As the XCAT phantom does not include the patient couch, we removed the patient couch from the CT patient volumes. For CBCT and MRI no couch is visible in the patient images.

The image intensities are windowed to the ranges given in Table \ref{table:ds_stats}. For CT and CBCT a fixed window was used. Since MRI intensities vary widely from image to image, the 10th and 90th percentile of each volume (whole 3D matrix) was used for windowing. For training, a linear intensity transformation is applied to transform the intensities from the windowing interval to [-1,1]. Normalization of training data is a crucial step in improving training performance, regardless of the normalization method used \cite{jacobsen2019analysis}.

\begin{table}[t]
\centering
\caption{Training data statistics. For CT and MRI the number of slices per image vary in the given interval.}
\label{table:ds_stats}
\setlength{\tabcolsep}{2pt}
\renewcommand{\arraystretch}{1.25}
\scalebox{0.8}{
\begin{tabular}{l|c|cccccc}
\hline
  & Parameter & Resolution (x/y/z) [mm] & Windowing & Volumes & Volume size & Arms & Age\\ 
\hline
CT Patient      & 100-130\,kVp & 1/1/2      & [-1024, 1500]\,HU    & 22  & $512\times512\times[52, 151]$ & no & 66 $\pm$ 9 \\
CT Phantom      & 90-120\,kVp & 1/1/2       & [-1024, 1500]\,HU     & 56   & $512\times512\times[80, 124]$ & no & 51 $\pm$ 14 \\
CBCT Patient    & 93-124.7\,kVp  & 0.486/0.486/0.486    & [-1024, 2000]\,HU     & 24    & $512\times512\times386$ & no & 67 $\pm$ 10 \\
CBCT Phantom    & 90-120\,kVp & 0.486/0.486/0.486      & [-1024, 2000]\,HU    & 56    & $512\times512\times386$ & no & 51 $\pm$ 14 \\
MRI Patient     & 3\,T & 1/1/3       & [10th , 90th]\,percentile     & 24  & $330\times450\times[48, 93]$ & yes & 67 $\pm$ 10 \\
MRI Phantom     & 3\,T & 1/1/3       & [10th , 90th]\,percentile    & 56  & $330\times450\times[59, 88]$ & yes & 51 $\pm$ 14 \\
\hline
\end{tabular}
}
\end{table}

\subsubsection{XCAT Phantom Data}
The XCAT model provides highly detailed whole-body anatomies. Organ masks can be easily obtained within the XCAT framework. Since CycleGANs maintain the geometry provided by the XCAT, the organ masks can be used as segmentation masks in the synthesized images. The phantom includes female and male models for varying ages. The heart beat and respiratory motions can be simulated and displacement fields of these motions can be generated. The anatomy and motion can be adapted by various parameters. This allows the creation of highly individual patient geometries. For the XCAT training data we generate one XCAT volume per XCAT model for each modality with 56 different models of varying ages. The XCATs include the whole liver and are generated with the same voxel spacing, windowing and normalization as the resampled patient data. Arms are included only in the MRI XCATs.

The XCAT phantom provides attenuation coefficients for all organs. We vary the simulated tube energy of the CBCT and CT phantoms from 90-120 keV in steps of 5 keV. This leads to a variation of attenuation coefficients in the phantoms. Afterwards, those are transformed into Hounsfield Units. To obtain CBCT and MRI XCAT data, we need to convert the CT XCAT. For the CBCT XCAT we apply a field of view mask obtained from the patient CBCTs, which is centered on the liver. For the MRI phantoms we replace the attenuation coefficients for each organ with simulated MRI values using the signal equation for the VIBE sequence. It ensures that the MRI signal is initialized with realistic values matching the MRI training data. This enables us to use the aforementioned intensity and gradient loss for the generation of synthetic MRI images, since the transformation with the CycleGAN is now monomodal. The signal intensities ($SI$) for the VIBE sequence in terms of acquisition parameters repetition time TR, echo time TE, and flip angle $\alpha$ and tissue-specific $T1$, $T2$ relaxation times, and proton density $\rho$ is given by:

\begin{align}
    SI = \frac{\rho \sin{\alpha} (1-\exp{-\frac{TR}{T1}})}{(1-\cos{\alpha}\exp{\frac{-TR}{T1}})}\exp{\frac{-TE}{T2}}.
\end{align}

We calculate the MRI intensity for all 44 abdominal organs present in the XCAT. The imaging parameters $TE = 4.54$\,ms, $TR = 7.25$\,ms, and $\alpha = 10^\circ$ are obtained from the patient VIBE scans. The values for the proton density $\rho$ are taken from \cite{wissmann2014mrxcat}. T1 and T2 relaxation times for 3\,T for blood and the spinal cord are obtained from \cite{stanisz2005t1} and the rest from \cite{de2004mr}. For organs with no available T1, T2 or $\rho$ we use values of similar organs. To simulate some organ variability, we randomly vary T1, T2, and $\rho$ by $\pm 5\,\%$ using a uniform distribution.

\subsection{Evaluation Metrics}

Quantification of the synthetic image quality is difficult, since there are no corresponding real images for comparison \cite{tmenova2019cyclegan}. Therefore, metrics that require a one-to-one correspondence like the mean absolute error (MAE) cannot be calculated between synthetic and real images. Instead, we calculate one-to-one corresponding metrics between the synthetic images and the XCATs, to investigate the magnitude of change from the XCAT phantoms. Real patient images and synthetic images are then compared by assessing their noise characteristics and voxel intensity distributions.

\subsubsection{Synthetic vs. XCAT}
The axial slices of the synthetic CT volumes are compared to the corresponding axial slices of the XCAT volumes with respect to anatomical accuracy. The MAE is calculated to assess the change of the intensity values. We exclude the background for the calculation of the MAE. The similarity of structure and features is evaluated using structural similarity index measure (SSIM) and feature similarity index measure (FSIM) \cite{Wang2004,Zhang2011}. Additionally, we calculate the edge preservation ratio (EPR) and edge generation ratio (EGR) \cite{Chen2016,russ2019synthesis}.

\subsubsection{Synthetic vs. Real Patient }

Regarding realistic noise characteristics and intensity distribution, the 3D synthetic volumes are compared to the 3D patient volumes.
For the noise characteristics, only liver voxels are considered. Limiting the noise considerations to the liver is reasonable, since the liver is a large and mostly homogeneous organ. We manually segmented the liver in 4 patients for each modality. The liver segmentations for the 56 synthetic images are provided by the XCAT phantom. The noise texture is evaluated using an estimation of the radial noise power spectrum (NPS). The radial NPS of the synthetic and patient images is compared by calculating the Pearson correlation coefficient, further called the NPS correlation coefficient (NCC) \cite{russ2019synthesis}. In addition to noise texture, we calculate the noise magnitude (NM), i.e. the standard deviation of the liver voxel intensities.

Furthermore, intensity distribution histograms of patient and synthetic images are calculated. To quantify their similarity, the Pearson correlation coefficient between them is calculated (HistCC).

\subsection{Proof of Principle Registration Evaluation}

We perform a proof of principle image registration to demonstrate the feasibility of the multimodal dataset for evaluation and thus development of registration algorithms. Our goal is to investigate different parameter settings to optimize the registration result. We implement the registration in Python 3.5 with SimpleITK 1.2.4. A non-rigid B-spline transform with a gradient descent optimizer, a learning rate of 1 and a maximum of 300 iterations is used. Three different registration metrics are considered, namely Mattes Mutual Information (MMI), Normalized Correlation (NC), and Mean Squares (MS). For the MMI, 50 histogram bins are used. The MS metric is only used for the monomodal CT to CT registration, since it is not suited for multimodal images. Additionally, we vary the spacing of the B-spline control points from 50\,mm to 150\,mm in steps of 20\,mm. For the multimodal and monomodal registrations this results in 12 and 18 different parameter settings, respectively.

The registration is performed on the synthetic data from all 56 XCAT models. We registered the CT, MRI and CBCT images in the inhaled state to the CT image in the exhaled state. To evaluate the registration, we take advantage of the liver organ masks obtained from the XCAT phantoms. The veins and arteries inside the liver are included in the liver mask by using a morphological closing operation. We apply the registration transform to the liver masks in the inhaled state and compare the result to the CT liver mask in the exhaled state. The overlap of the two masks is assessed by calculating the Dice similarity coefficient (DSC).

\section{Results}

\subsection{Synthetic Images}

First, we consider the metrics that compare the synthetic images with the XCAT phantoms shown in the upper half of Table \ref{tab:synthetic_eval}. The FSIM and SSIM indicate that image structures and features were well preserved in the CT and CBCT images, whereas the synthetic MRIs showed little structural and feature similarity to the XCATs. Regarding edges, the EPR is similar for all modalities, whereas the EGR is largest for the CBCT images. The MAE is slightly larger than the NM (synthetic) for every modality. The MAE for CBCT is more than twice as high as the MAE for CT.

\begin{table}[ht]
	\centering
	\caption{Image quality metrics for the evaluation of the synthetic images.}
	\label{tab:synthetic_eval}
	\begin{tabular}{c|ccc}
		\hline
		                & CBCT   & CT  & MRI  \\
\hline
SSIM                    & 0.85 $\pm$ 0.05  & 0.94 $\pm$ 0.02 & 0.59 $\pm$ 0.04  \\ 
FSIM                    & 0.82 $\pm$ 0.03  & 0.82 $\pm$ 0.02 & 0.51 $\pm$ 0.02  \\
EPR                     & 0.47 $\pm$ 0.06  & 0.43 $\pm$ 0.04  & 0.40 $\pm$ 0.03  \\
EGR                     & 3.0 $\pm$ 0.7  & 1.9 $\pm$ 0.3  & 1.7 $\pm$ 0.2  \\
MAE                     & 109 $\pm$ 14   & 51 $\pm$ 16    & 37 $\pm$ 6 \\ \hline
NCC    & 0.997 $\pm$ 0.001  & 0.980 $\pm$ 0.010  & 0.86 $\pm$ 0.04 \\ 
NM (Synthetic) & 52 $\pm$ 13    & 39 $\pm$ 5    & 25 $\pm$ 3 \\
NM (Patient)   & 60 $\pm$ 16 & 39 $\pm$ 19  & 22 $\pm$ 5  \\
HistCC                  & 0.994 $\pm$ 0.003    & 0.999 $\pm$ 0.002    & 0.94 $\pm$ 0.03 \\ \hline

	\end{tabular}
\end{table}

\begin{figure}[t]
\centering
    \begin{tabular}{C{0.32\textwidth}C{0.32\textwidth}C{0.32\textwidth}}
        \includegraphics[width=0.32\textwidth]{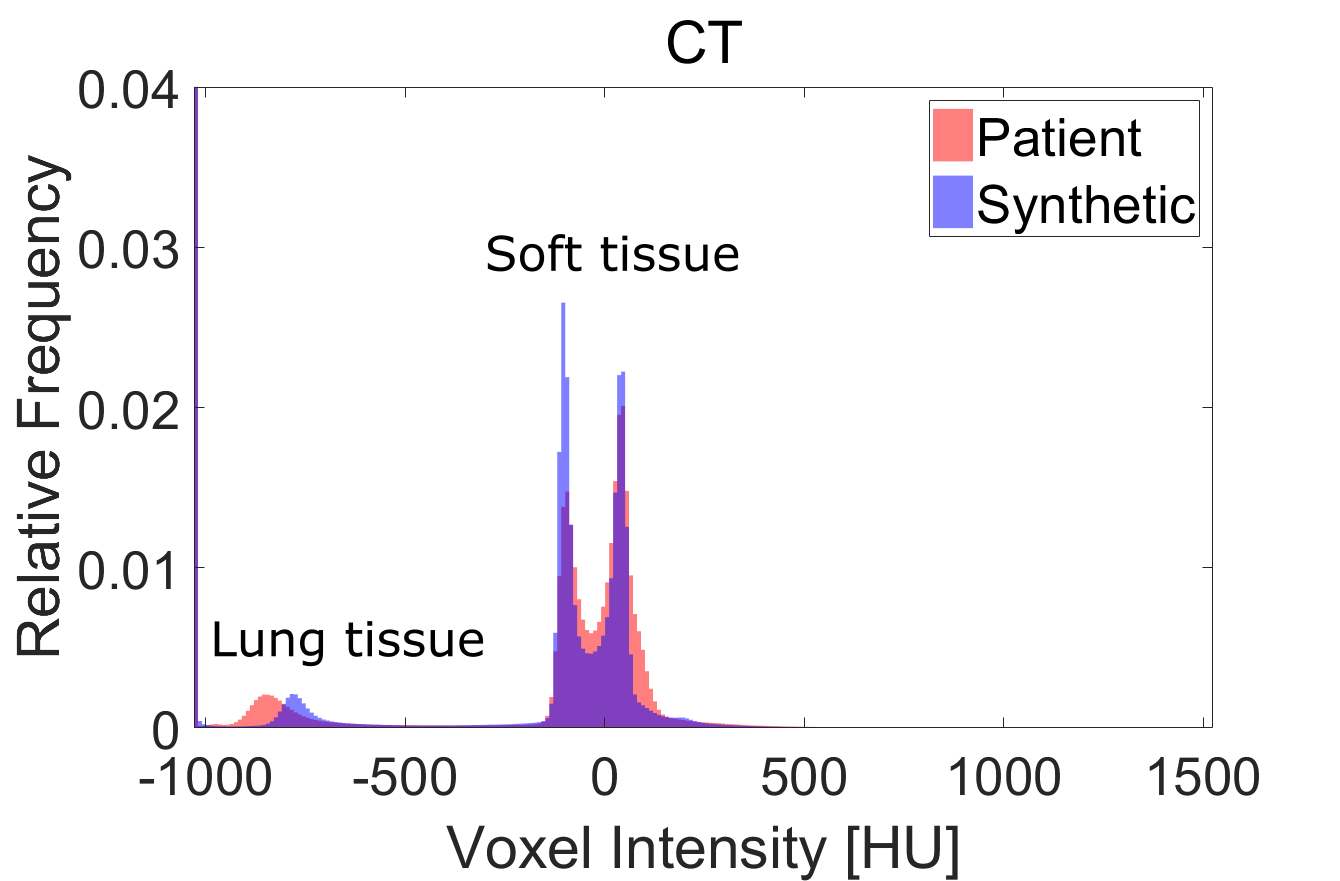} &
        \includegraphics[width=0.32\textwidth]{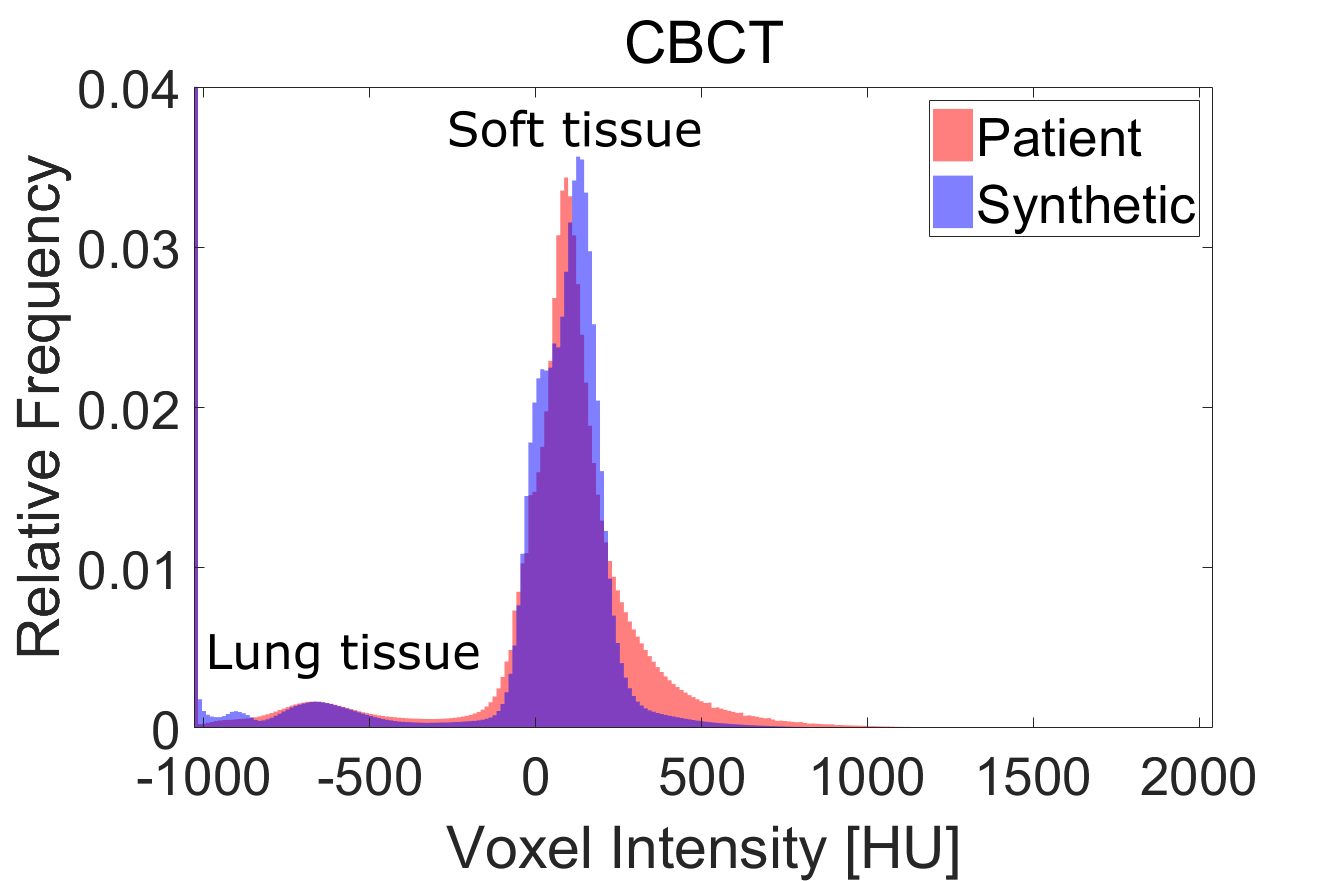} &
        \includegraphics[width=0.32\textwidth]{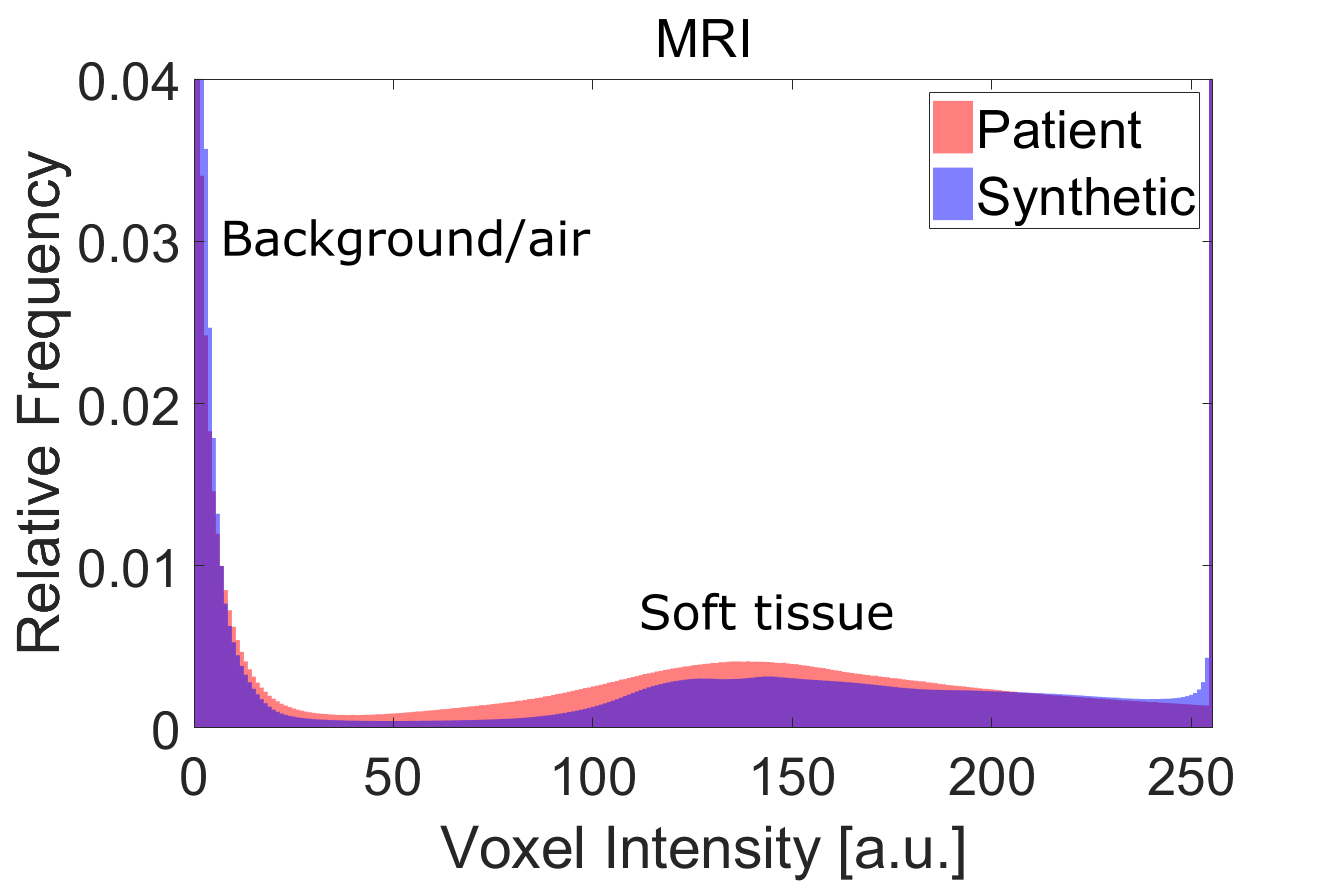} \\
    \end{tabular}
	\caption{Intensity histograms of the patient and synthetic images averaged over all volumes. Note that the background peaks are cropped.}
	\label{fig:evaluation_histo}
\end{figure}

Secondly, we compare the synthetic images to the patient images. The two right columns of Fig. \ref{fig:synthetic_result} show axial synthetic and patient slices of each modality. Qualitatively, the style of the synthesized images is in good agreement with the real patient images. To quantify this observation we compared the noise characteristics and voxel intensity distribution of the synthetic images to the patient images, the results are listed in the lower half of Table \ref{tab:synthetic_eval}. A high NCC for all modalities indicates that the noise texture was emulated realistically, albeit the NCC is slightly smaller for the synthetic MRI images. For all modalities, the NM (synthetic) is in excellent agreement with the NM (patient). In Fig. \ref{fig:evaluation_histo} the intensity histograms are shown. In general, the synthetic intensity distributions match the patient intensity distributions nicely. This is underlined by the overall high HistCC values in Table \ref{tab:synthetic_eval}. However, for CT and CBCT the soft tissue peaks are modeled a bit too narrowly. The lung tissue peak is shifted towards higher CT numbers for the CT. In the MRI, the soft tissues is slightly underrepresented.

\subsection{Proof of Principle Registration}

\begin{figure}[t]
\centering
    \begin{tabular}{C{0.31\textwidth}C{0.31\textwidth}C{0.31\textwidth}}
        \includegraphics[width=0.31\textwidth]{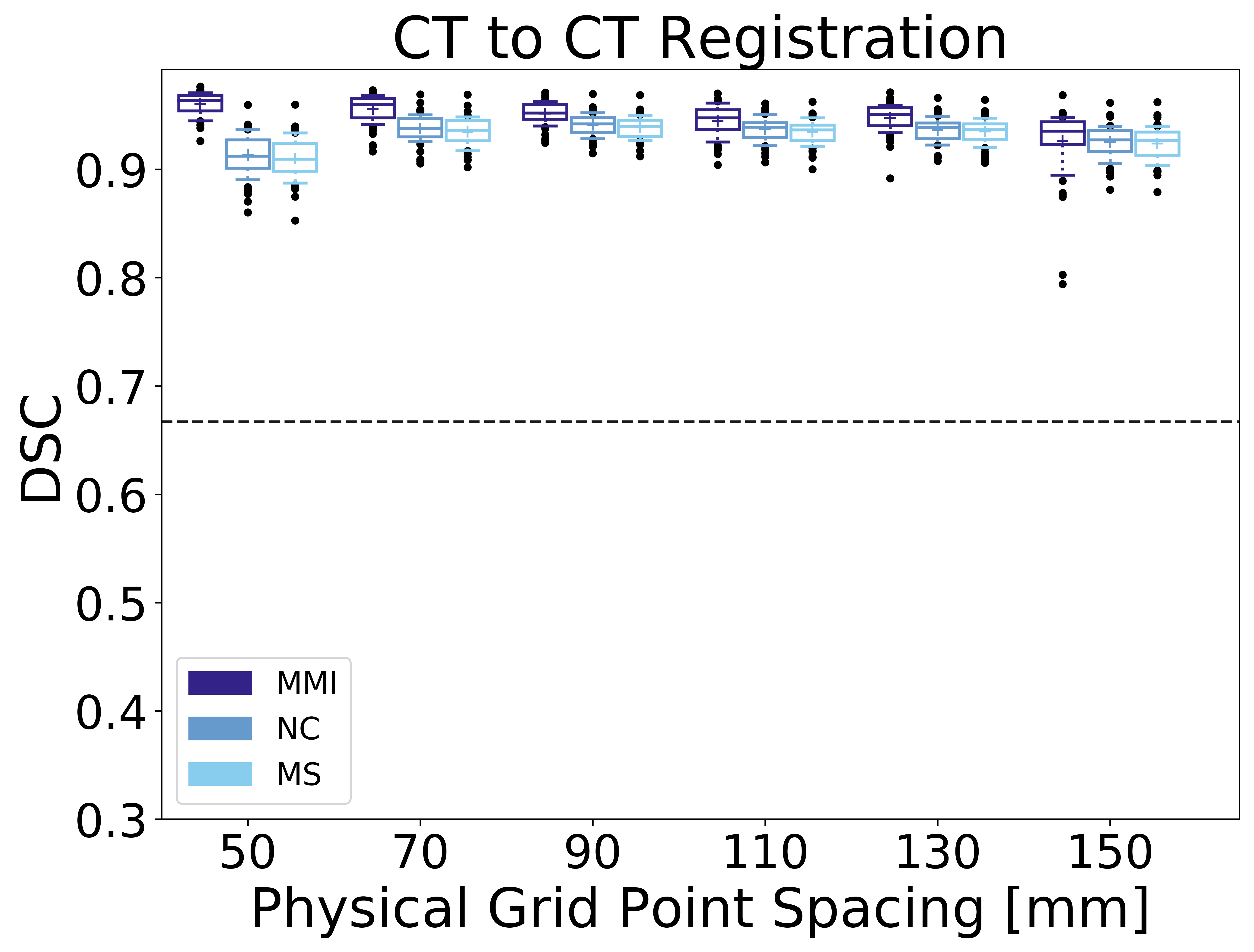} &
        \includegraphics[width=0.31\textwidth]{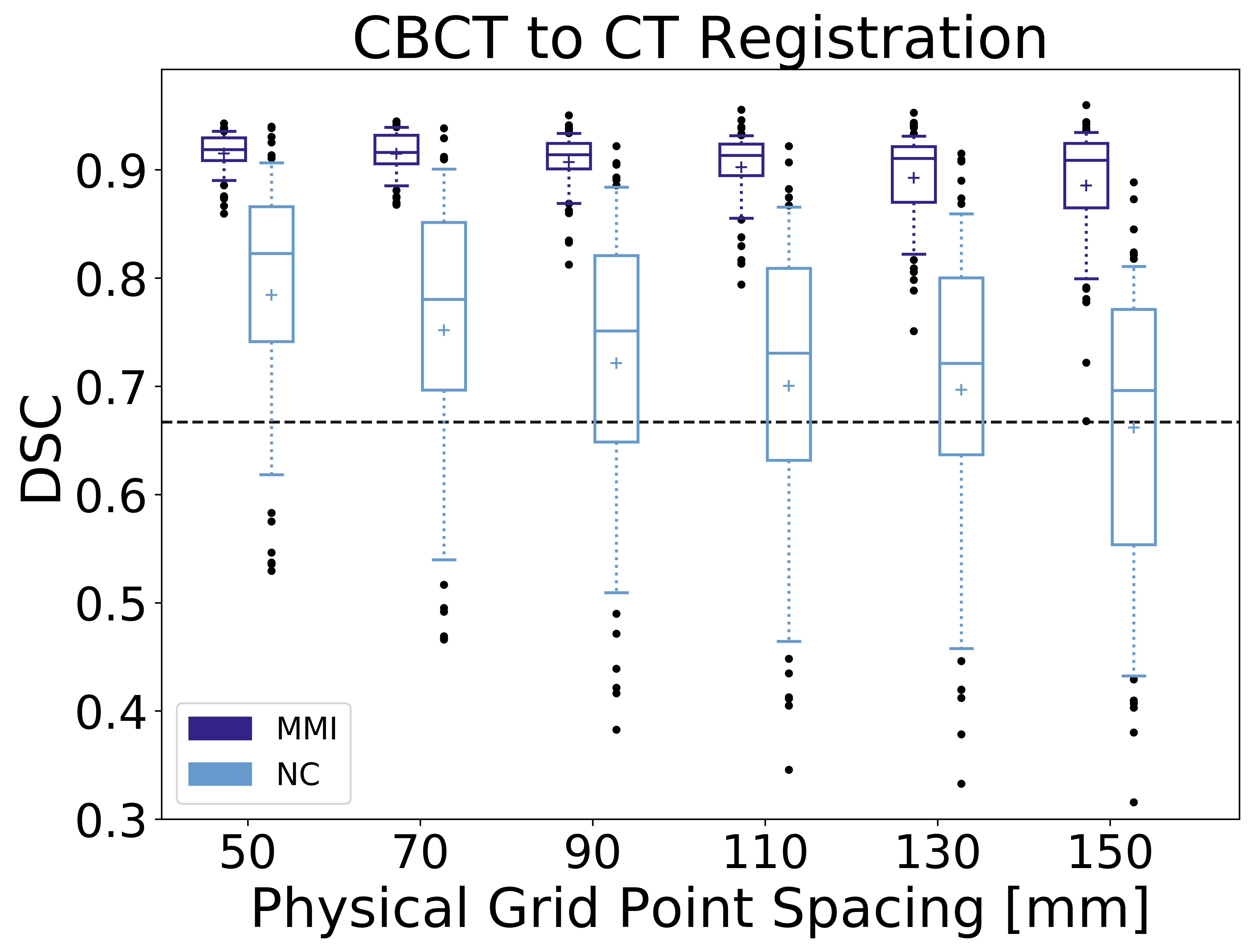} &
        \includegraphics[width=0.31\textwidth]{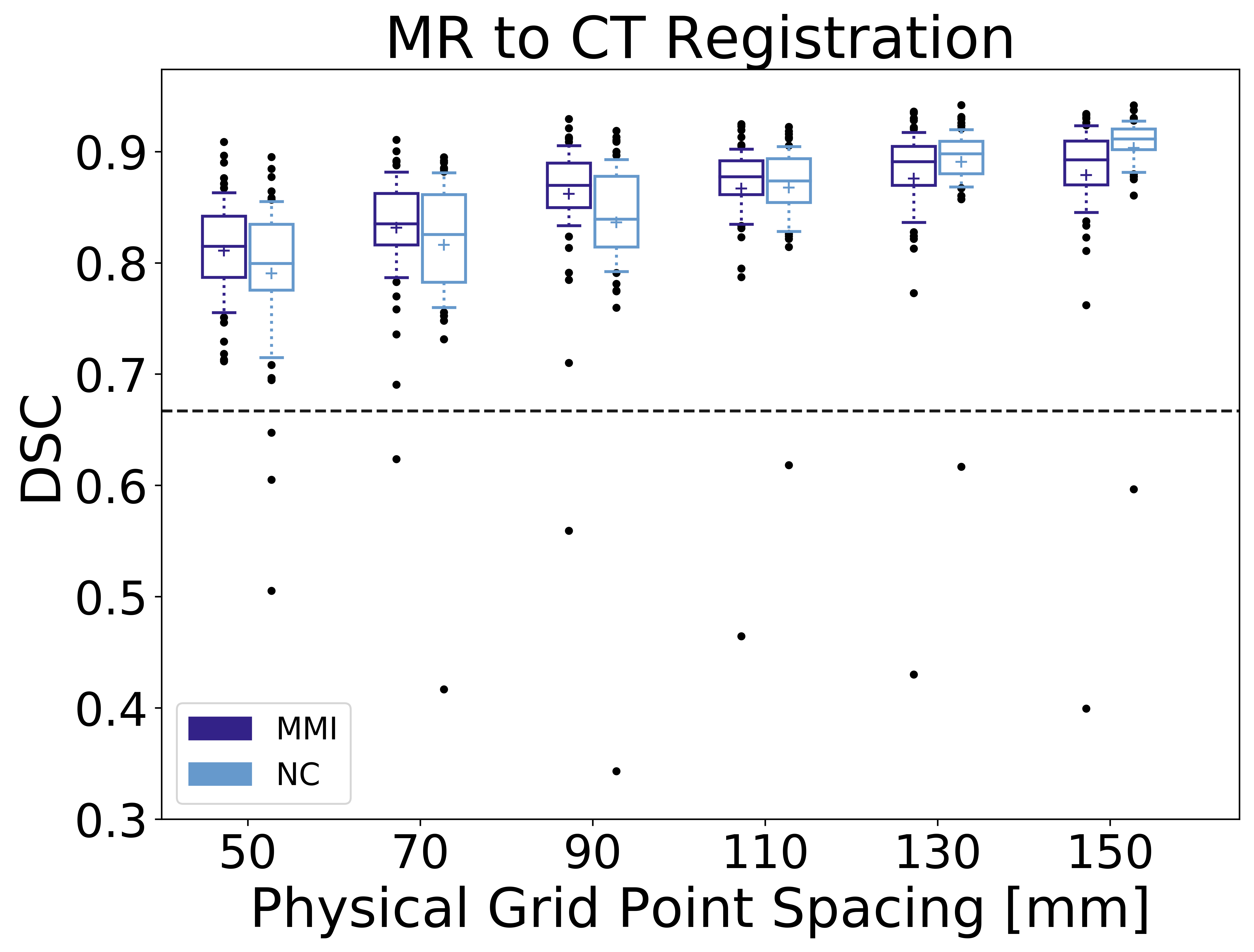} \\
    \end{tabular}
    
	\caption{DSC for the proof of principle registrations with 56 data points each. The mean is marked as a "+" and the whiskers indicate the 10th and 90th percentile. The dashed horizontal line shows the mean pre-registration DSC.}
	\label{fig:registration_boxplot}
\end{figure}

The DSC for the evaluation of the proof of principle registration is shown in Fig. \ref{fig:registration_boxplot}. The monomodal CT to CT registration yielded good results for all registration metrics and grid point spacings, with the best result for MMI with 50\,mm grid point spacing. For CBCT, the MMI again worked well, whereas the registrations using the NC mostly failed. The best results were again obtained with MMI and a grid point spacing of 50\,mm. For MRI, the registrations with MMI and NC yielded similar results with the best result obtained for NC with a grid point spacing of 150\,mm. Overall, the monomodal CT to CT registration achieved the best results.

\begin{figure}[t]
\centering
\setlength{\tabcolsep}{2px}
\renewcommand{\arraystretch}{1}
    \begin{tabular}{m{0.3cm}C{0.30\textwidth}C{0.30\textwidth}C{0.30\textwidth}}
        & \large Pre-Registration & \large Post-Registration  &  \large Ground Truth Image \\
        
        \rotatebox[]{90}{\large CBCT} &
        \includegraphics[width=0.30\textwidth]{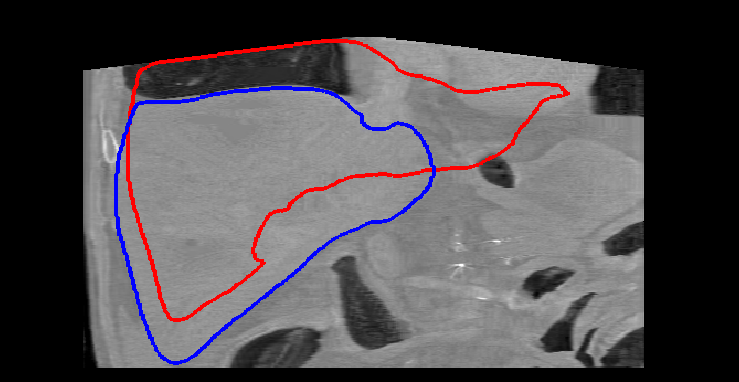} &
        \includegraphics[trim = 0px 10px 0px 10px, clip, width=0.30\textwidth]{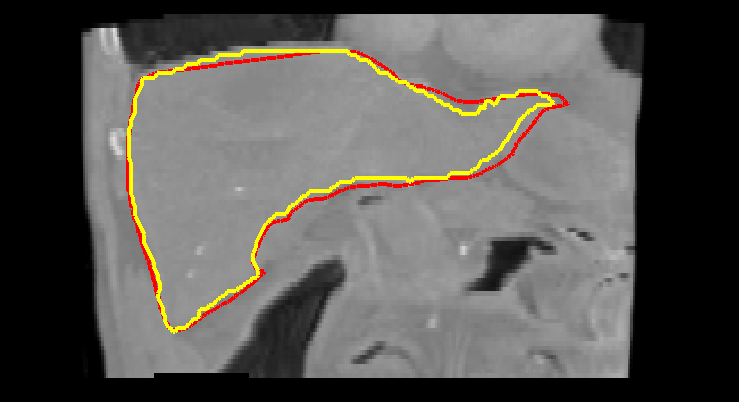} &
        \includegraphics[trim = 0px 10px 0px 10px, clip, width=0.30\textwidth]{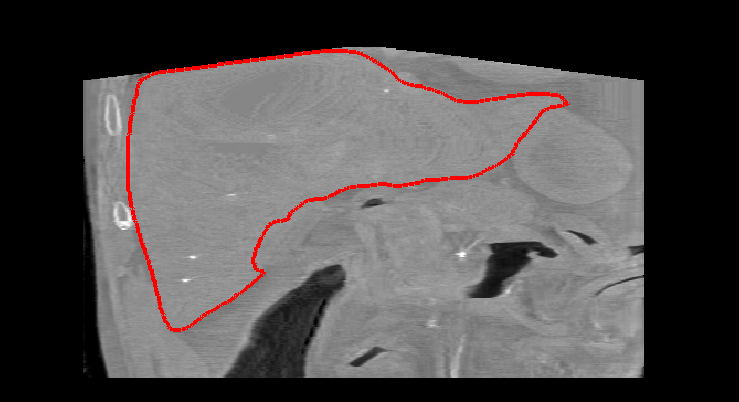} \\
        
        \rotatebox[]{90}{\large CT} &
        \includegraphics[trim = 0px 1px 0px 1px, clip, width=0.30\textwidth]{CT3_Mask_fix.png} &
        \includegraphics[trim = 0px 1px 0px 1px, clip, width=0.30\textwidth]{CT_Trafo_Mask_fix.png} &
        \includegraphics[trim = 0px 1px 0px 1px, clip, width=0.30\textwidth]{CT1_Mask_fix.png} \\
        
        \rotatebox[]{90}{\large MRI} &
        \includegraphics[trim = 0px 10px 0px 10px, clip, width=0.30\textwidth]{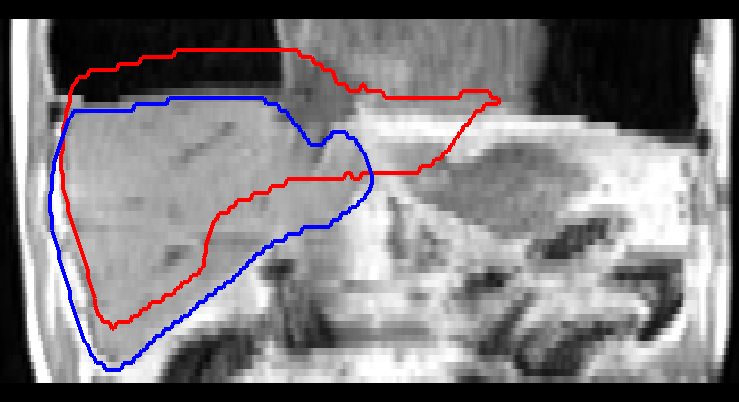} &
        \includegraphics[trim = 0px 10px 0px 10px, clip, width=0.30\textwidth]{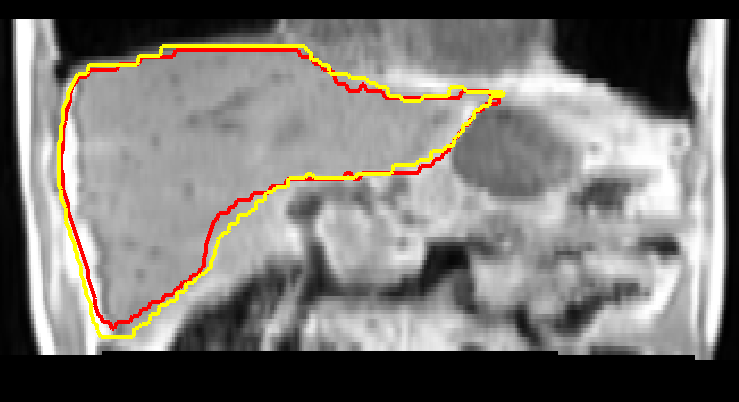} &
        \includegraphics[trim = 0px 10px 0px 10px, clip, width=0.30\textwidth]{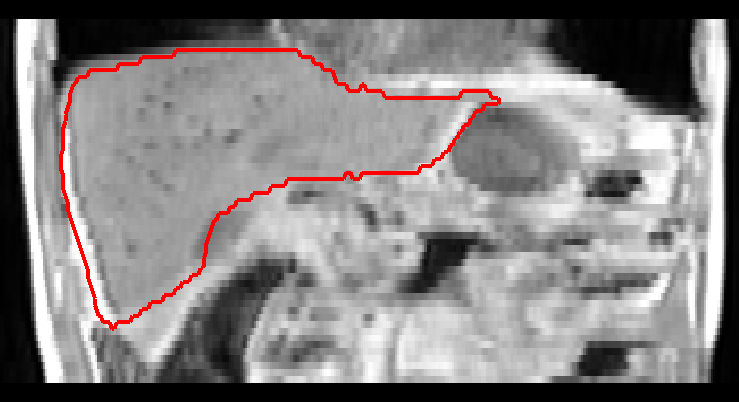} \\
        
    \end{tabular}
\caption{Pre- and Post-registered images and the corresponding ground truth. Red contours indicate the ground-truth boundaries of the liver (target). Blue and yellow contours represent the boundaries of the liver before and after deformation respectively.}
\label{fig:registered_images}
\end{figure}

Coronal views of the registration results for the best settings of each modality are visualized in Fig. \ref{fig:registered_images}. The registered images in the middle row show a large similarity to the ground truth. This observation is further supported by the overlaid liver contours. The post-registration liver contour (yellow) is in high agreement with the ground truth liver contour (red).

\section{Discussion}

The image quality metrics in Table \ref{tab:synthetic_eval} demonstrate that our framework provides reaslistic multimodal image data. Low SSIM and FSIM for the MRI images indicate, that image values of the homogeneous organs in the MRI XCAT phantoms needed to be altered more strongly by the networks in comparison to the CT XCATs. The lower SSIM, FSIM, and EPR for MRI are likely a result of lower weighting of the gradient loss and intensity loss.

The ratio of MAE to NM (synthetic) is 2.1, 1.3, and 1.5 for CBCT, CT, and MR, respectively. Assuming normally distributed noise, the ratio of MAE to NM is only about 0.8 \cite{geary1935ratio}. This means that the MAE cannot be attributed to noise alone. The large MAE for the CBCT images compared to the CT images could be due to the introduction of metal artifacts, as the patient CBCTs showed metal artifacts in the liver caused by medical instruments. This is supported by a large EGR for CBCT.

For all modalities realistic noise texture and magnitude was achieved. Additionally, the voxel intensity distribution was modeled adequately. Most of the discrepancies between the patient and synthetic histograms in Fig. \ref{fig:evaluation_histo} can be explained by inspecting the XCAT phantoms. The deviation of the CT lung peaks (synthetic \mbox{-780\,HU}, patient \mbox{-835\,HU}) can be explained by an overestimated initial lung value of \mbox{-760\,HU} given by the XCAT. The narrow soft tissue peaks for CT and CBCT could be due to insufficient variation in organ attenuation coefficients. The under representation of soft tissue in the synthetic MRI is due to the body size of the patients and XCATs. We found that in the MRI patient dataset 66.5\,\% of the image voxels show the body, whereas for the MRI XCAT dataset, it is only 46.5\,\%. A rather large HistCC of \mbox{0.94 $\pm$ 0.03} was still achieved, since this under representation has only a minor effect on the correlation between the histograms. We prepared the XCAT data such that it matches the patient dataset as good as possible, see Table \ref{table:ds_stats}. In the future we will consider the patient body size beforehand and adjust the XCAT body size accordingly.

An important requirement for using synthetic data to evaluate or train registration algorithms is a realistic respiratory deformation model. The XCAT framework allows the respiratory rate, the amount of diaphragmatic motion, the amount of chest expansion, and the amount of cardiac motion due to respiration to be varied. In addition, curves controlling the diaphragm motion and chest expansion can be individualized. However, as the breathing patterns of patients are complex and diverse, the XCAT model may still lack generalizability.

The results of the proof of principle registration demonstrate that the synthetic dataset can be used to evaluate different registration algorithms. We were able to evaluate the performance of different registration algorithms and to fine tune parameter settings. It is noticeable that the registration for CT and CBCT works better for small B-spline grid point spacings, while it is the opposite for MRI. Further studies are needed to assess whether this is due to the smaller image size of the MRI data and thus the total number of B-spline control points or to other imaging characteristics of MRI and CT. Computation time can be measured and taken into consideration. For example, registrations with smaller grid point spacings take much longer. Thus, choosing MMI with 150\,mm for CT and CBCT registrations might be a reasonable trade-off, as the registration quality is only slightly lower, while the registration time is substantially reduced. The availability of organ masks enabled a rather simple registration evaluation.

\section{Conclusion}

The presented simulation framework can be used to extend small datasets by transferring the style of the dataset onto the geometry given by the XCAT phantom. The obtained datasets can serve as a ground truth for image registration.
A multimodal dataset consisting of T1-weighted MRI, CT and CBCT images was created and used to demonstrate the refinement and evaluation of multimodal image registration algorithms.
In the future, the framework will be extended to other modalities, such as T2-weighted MRI or PET, which can further boost the performance of multimodal methods. An extension to other body regions, such as the thorax or pelvis, is also possible. Synthetic images over larger body regions are especially interesting for whole body segmentation. Expansion of datasets using this method provides a promising tool to overcome the dearth of medical training data.

\begin{acknowledgements}
We gratefully acknowledge the support of NVIDIA Corporation with the donation of the NVIDIA Titan Xp GPU used for this research.\\
\textbf{Funding} This research project is part of the Research Campus M$^2$OLIE and funded by the German Federal Ministry of Education and Research (BMBF) within the Framework 'Forschungscampus - Public-Private Partnership for Innovation' under the funding code 13GW0388A.\\
\textbf{Conflict of Interest} The authors declare that they have no conflict of interest.\\
\textbf{Human and animal rights} We are in compliance with ethical standards.\\
\textbf{Ethics approval} Ethical approval for this study was obtained from the ethics committee of the medical faculty Mannheim (2016-863R-MA).
\end{acknowledgements}

\bibliographystyle{spmpsci}      
\bibliography{library}   

\end{document}